\documentclass[aps, prl, twocolumn, preprintnumbers, nofootinbib, superscriptaddress]{revtex4-1}

\usepackage{amsmath,amssymb}
\usepackage[usenames,dvipsnames,svgnames,table]{xcolor}
\usepackage[colorlinks,citecolor=DarkGreen,linkcolor=FireBrick,urlcolor=FireBrick,linktocpage]{hyperref}
\usepackage{graphicx}
\usepackage{times}

\def\bC{\mathbb{C}}
\def\bZ{\mathbb{Z}}
\def\SL{\mathop{\mathrm{SL}}}
\def\Arf{\mathop{\mathrm{Arf}}}
\def\Arg{\mathop{\mathrm{Arg}}}
\def\Mp{\mathop{\mathrm{Mp}}}
\def\U{\mathrm{U}}

\begin{document}

\title{Anomaly of the Electromagnetic Duality of Maxwell Theory}
\preprint{IPMU-19-0068}
\preprint{TU-1088}

\author{Chang-Tse Hsieh}
\affiliation{Kavli Institute for the Physics and Mathematics of the Universe (WPI), \\
 University of Tokyo,  Kashiwa, Chiba 277-8583, Japan}
\affiliation{Institute for Solid State Physics, University of Tokyo, Kashiwa, Chiba 277-8581, Japan}
\author{Yuji Tachikawa}
\affiliation{Kavli Institute for the Physics and Mathematics of the Universe (WPI), \\
 University of Tokyo,  Kashiwa, Chiba 277-8583, Japan}
\author{Kazuya Yonekura}
\affiliation{ Department of Physics, Tohoku University, Sendai 980-8578, Japan
%Particle Theory and Cosmology Group, Department of Physics, Graduate School of Science, Tohoku University, Aoba-ku, Sendai 980-8578, Japan
}

%\date{\today}

%-----------------------------------------
\begin{abstract}
We consider the ($3{+}1$)-dimensional Maxwell theory in the situation where going around nontrivial paths in the spacetime involves the action of the duality transformation exchanging the electric field and the magnetic field, as well as its $\SL(2,\bZ)$ generalizations.
We find that the anomaly of this system in a particular formulation is 56 times that of a Weyl fermion.
This result is derived in two independent ways: 
one is by using the bulk  symmetry protected topological phase in $4{+}1$ dimensions characterizing the anomaly,
and the other is by considering the properties of a ($5{+}1$)-dimensional superconformal field theory known as the E-string theory.
This anomaly of the Maxwell theory plays an important role in the consistency of string theory.
\end{abstract}

\pacs{}
\maketitle

\section{Introduction}

Every physicist knows that the electromagnetic field is described classically by the Maxwell equation,
and that it is invariant under the electromagnetic duality $S:(\mathbf{E},\mathbf{B}) \mapsto (\mathbf{B},-\mathbf{E})$.\footnote{%
For some of the early contributions to the study of the duality transformation, see e.g.,~\cite{Deser:1976iy,Deser:1981fr}.
}
The properties of the electromagnetic duality in quantum theory might not be as well known to physicists in general and, in fact, are not very well understood in the literature.
This is particularly true when going around a nontrivial path in the spacetime results in a duality transformation.\footnote{%
One example is a periodic boundary condition twisted by duality:
$\mathbf{E}(x+L,y,z)=\mathbf{B}(x,y,z)$ and $\mathbf{B}(x+L,y,z)=-\mathbf{E}(x,y,z)$.
This particular setup was studied by O.~Ganor and his collaborators \cite{Ganor:2008hd,Ganor:2010md,Ganor:2012mu,Ganor:2014pha},
but what happens in a more general situation remains unanswered,
to the authors' knowledge.
There is also a series of interesting papers on the flux sectors of the Maxwell theory by G.~W.~Moore and his collaborators \cite{Freed:2006ya,Freed:2006yc,Kitaev:2007ed},
which are related to the inherent  self-dual nature of the Maxwell theory.
Another intriguing scenario is to consider a Maxwell theory with dynamical ``duality gauge fields'', which might be thought of as a generalization of the Alice electrodynamics \cite{Schwarz:1982, Bucher:1992} where the charge conjugation $C=S^2$ is gauged.
}
In this Letter, we focus on a feature of the Maxwell theory and its duality symmetry in such a situation, namely the fact that it has a quantum anomaly \cite{Kravec:2013pua,Seiberg:2018ntt}, which we explicitly determine.

We recall that a quantum theory in $d{+}1$ dimensions with a symmetry group $G$ can have a quantum anomaly, in the sense that its partition function has a controllable phase ambiguity.
Our modern understanding is that such a theory is better thought of as living on the boundary of a symmetry protected topological (SPT) phase in the $[(d{+}1){+}1]$-dimensional bulk.
It was noticed in the last few years in \cite{Wang:2013zja,Kravec:2013pua,Kravec:2014aza,Thorngren:2014pza,Wang:2018qoy} that 
a version of the Maxwell theory (often called the all-fermion electrodynamics, where all particles of odd charge are fermions) has a global gravitational anomaly and lives on the boundary of a certain bulk SPT phase.
As we will see, this result is a special case of the anomaly and the corresponding bulk SPT phase that we find for the duality symmetry.

We study the anomaly and the bulk SPT phase by imitating the relationship between the $(1{+}1)$-dimensional chiral boson and the $(2{+}1)$-dimensional $\U(1)_1$ Chern-Simons theory and its generalization to the $[(4n{+}1){+}1]$-dimensional self-dual form field and the $[(4n{+}2){+}1]$-dimensional bulk theory studied, e.g.,~in \cite{Gukov:2004id,Belov:2006jd,Monnier:2016jlo,Monnier:2017klz,Monnier:2017oqd,Monnier:2018nfs,Monnier:2018cfa}.
The essential point is that the $(3{+}1)$-dimensional Maxwell theory with a nontrivial background for its duality symmetry is a self-dual field, and we can utilize the techniques developed in the papers listed above to study it.
One of our main messages is that the subtle and interesting issues concerning the self-dual fields studied in the past already manifest themselves in the case of the Maxwell theory once the non-trivial background for its duality symmetry is turned on.

Before proceeding, we note that the electromagnetic duality group in the quantum theory is, in fact, the 2-dimensional special linear group $\mathrm{SL}(2,\mathbb{Z})$ over the integers acting on the lattice $\bZ^2$ of the electric and magnetic charges.
Its effect on the Maxwell theory on a curved manifold was carefully analyzed in \cite{Witten:1995gf, Verlinde:1995em} and it was interpreted as a mixed $\SL(2,\bZ)$-gravitational anomaly in  \cite{Seiberg:2018ntt}.
Our result in this Letter can be considered as the determination of the pure $\SL(2,\bZ)$ part of the anomaly.

Our computation shows that the anomaly of the Maxwell theory is 56 times that of a Weyl fermion in a certain precise formulation of the duality.
Where does this number 56 come from?
We will provide an answer using the property of a $(5{+}1)$-dimensional superconformal field theory originally found in \cite{Ganor:1996mu,Seiberg:1996vs}
and known as the E-string theory;
the name comes from the fact that it has $E_8$ global symmetry.
The E-string theory has two branches of vacua, called the tensor branch and the Higgs branch. 
On the Higgs branch the $E_8$ symmetry is Higgsed to $E_7$, which acts on 28 fermions via its 56-dimensional fundamental representation; this is possible since
a pseudo-real representation $R$ with $\dim R = 2k$ can act on $k$ fermions in $5{+}1$ dimensions because the spin representation $S$ in $5{+}1$ dimensions is pseudo-real
and we can impose the Majorana condition on $R \otimes S$.
When one moves to the tensor branch, the $E_8$ symmetry is restored and a self-dual tensor field appears.
By compactifying this system on $T^2$, one finds that one Maxwell field is continuously connected to 56 Weyl fermions, showing that they should have the same anomaly.
The electromagnetic duality is formulated as the $\SL(2,\bZ)$ acting on this torus $T^2$, 
and therefore is geometrized in this formulation.
This means that both the purely $\SL(2,\bZ)$ part and the mixed gravitational-$\SL(2,\bZ)$ part of the $(3{+}1)$-dimensional anomaly come from the purely gravitational anomaly of the $(5{+}1)$-dimensional theory.
These statements about the anomaly are valid if the $E_8$ background field is turned off.

The rest of the Letter is organized as follows.
We start by recalling how the anomaly of a $(1{+}1)$-dimensional chiral boson is captured by the phase of the partition function of the $(2{+}1)$-dimensional U(1) Chern-Simons theory at level 1.
We outline the path integral computation of its phase, 
as well as how this can be matched with the anomaly of a $(1{+}1)$-dimensional chiral fermion.
We then adapt this discussion to the anomaly of the $(3{+}1)$-dimensional Maxwell theory and the corresponding $(4{+}1)$-dimensional bulk $BdC$ theory.
We will see that the anomaly computed in this way reproduces the known anomaly when the $\SL(2,\bZ)$ background is trivial.
We then consider the case of nontrivial $\SL(2,\bZ)$ backgrounds on $S^5/\bZ_k$, for $k=2,3,4$, and $6$,
and we note that the resulting phase is equal to 56 times that of a charged Weyl fermion. This plays an important role in the consistency of the O$3^-$-plane and its generalizations.
Finally, we explain why the anomaly of the Maxwell theory has to be 56 times that of a charged Weyl fermion, in terms of the six-dimensional superconformal field theory known as the E-string theory.
More details will be provided in a longer version of the Letter \cite{CTYtoappear}.

\section{Warm-up: Anomaly of $(1{+}1)$-dimensional chiral boson in terms of $(2{+}1)$-dimensional U(1) Chern-Simons}
We start by recalling the well-understood case of the anomaly of the $(1{+}1)$-dimensional chiral boson at the free fermion radius.
This theory naturally lives at the boundary of the $(2{+}1)$-dimensional U(1) Chern-Simons theory at level $k=1$, for which the  Euclidean action is $-S_{k=1}=\pi i\int (A/2\pi) (F/2\pi)$ \cite{Witten:1988hf,Bos:1989wa,Gukov:2004id}.
The anomaly is then characterized by the partition function of this Chern-Simons theory on closed 3-dimensional manifolds $M_3$.

Let us recall that the action at level 2, $-S_{k=2}=2\pi i\int (A/2\pi) (F/2\pi)$, is well-defined modulo $2\pi i$ when the manifold is oriented.
However, there is a problem in dividing it by two.
To make the action $S_{k=1}$ well-defined modulo $2\pi i$, it is known that we need to pick a spin structure  \cite{Belov:2005ze}.
Once this is done, the path integral can be performed explicitly, because the theory is free.
The details are given e.g.~in \cite{Witten:1988hf,Manoliu:1996fx,Jeffrey:2009vp,JeffreyBrendan2}. %,Guadagnini:2013sb,Guadagnini:2014mja}.
Very roughly, we split the gauge field $A$ into a sum of the flat but topologically nontrivial part and the topologically trivial but non-flat part.
Assuming, for simplicity, that flat connections on $M_3$ are isolated, we have \begin{multline}
 Z_\text{U(1)CS}(M_3)=\left[ \int [DA]_\text{top.trivial} e^{\pi i\int (A/2\pi) (F/2\pi)} \right]\\
\times  \left[\sum_{A:\text{flat}} e^{\pi i\int (A/2\pi) (F/2\pi)}\right].
\end{multline} 
Let us rewrite its phase.

The phase of the first term can be written in terms of the eta invariant of the signature operator $*d$: \begin{equation}
\tfrac1{2\pi}\Arg \int [DA]_\text{top.trivial} e^{\pi i\int (A/2\pi) (F/2\pi)} = -\tfrac18 \eta_\text{signature}. 
\end{equation}
Here and below, the equality of the phase is modulo one and is simply denoted by the equal sign ($=$).
The phase of the second term can be rewritten as \begin{equation}
\tfrac1{2\pi}\Arg  \sum_{c\in H^2(M_3,\bZ)} q(c)  =:  \Arf(q) 
\end{equation}
where $c=c_1(F)$ is the first Chern class of the gauge bundle and $q(c):=e^{\pi i\int (A/2\pi) (F/2\pi)}$.

We note that $q(c)$ is simply the exponentiated level-1 classical action  evaluated at a flat $A$.
As recalled above, defining it requires something more than an oriented manifold and the integration on it. 
Mathematically, $q$ is known as a quadratic refinement of the torsion pairing on $H^2(M_3,\bZ)$.
The Arf invariant $\Arf(q)$ is defined by the equation above and is known to take values in one eighth of an integer.
We end up with the formula \begin{equation}
\tfrac1{2\pi}\Arg Z_\text{U(1)CS}(M_3)= -\tfrac18\eta_\text{signature} +  \Arf(q).
\label{formula}
\end{equation}

Let us now recall that a chiral boson can be fermionized. 
Then the bulk theory can be taken to be the $(2{+}1)$-dimensional fermion with infinite mass, for which the partition function has the phase \cite{AlvarezGaume:1984nf} \begin{equation}
\tfrac1{2\pi}\Arg Z_\text{fermion}(M_3) = \eta_\text{fermion}.
\end{equation}
The values of $\eta_\text{signature}$ and $\eta_\text{fermion}$ on lens spaces are known in the literature, e.g., \cite{GilkeyOdd}.
For example, on $M_3=S^3/\bZ_2$,  $\eta_\text{signature}=0$, whereas $\Arf(q)$ and $\eta_\text{fermion}$ can be either $1/8$ or $-1/8$, depending on the spin structure.
On $M_3=S^3/\bZ_3$, 
$\eta_\text{signature}=2/9$, $\Arf(q)=1/4$, and $\eta_\text{fermion}=2/9$, 
as there is a unique spin structure.
We indeed confirm \begin{equation}
 -\tfrac18\eta_\text{signature} +  \Arf(q) = \eta_\text{fermion},
 \label{bosefermi}
\end{equation}
which can be proved using a mathematical result \cite{BrumfielMorgan}.
We note that $\eta_\text{signature}$ is independent of the spin structure
but $\Arf(q)$ does depend on the spin structure.
In other words, the spin structure provides us  the quadratic refinement.

\section{The anomaly of the Maxwell theory}
The analysis of the anomaly of the $(1{+}1)$-dimensional chiral boson we recalled above was generalized to the $[(4n{-}3){+}1]$-dimensional self-dual form fields in \cite{AlvarezGaume:1983ig} at the perturbative level. 
The study of the corresponding $[(4n{-}2){+}1]$-dimensional theory in the bulk, generalizing the $(2{+}1)$-dimensional Chern-Simons theory, was carried out in detail in \cite{Belov:2006jd,Monnier:2016jlo,Monnier:2017klz,Monnier:2017oqd,Monnier:2018nfs,Monnier:2018cfa}.
The bulk theory has the action $-S=\pi i \int (A/2\pi)d(A/2\pi)$, where $A$ is now a $(2n{-}1)$-form gauge field.
Assuming $H^{2n-1}(M_{4n-1},{\mathbb R})=0$, the phase of the partition function still has the form of Eq.~\eqref{formula},
where $q$ is now a quadratic refinement of the torsion pairing on $H^{2n}(M_{4n-1},\bZ)$,
and its choice is not obviously related to the choice of the spin structure.

Here, we are more interested in the $(3{+}1)$-dimensional Maxwell theory. 
The natural generalization in this case is to consider the bulk theory with the action $-S=\pi i \int [ (B/2\pi) d (C/2\pi) - (C/2\pi) d (B/2\pi)]$, where $B$ and $C$ are two 2-form gauge fields to be path-integrated over \cite{Kravec:2013pua}.
This action has the $\SL(2,\bZ)$ symmetry acting on $(B,C)$, which corresponds to the duality symmetry of the Maxwell theory \cite{Kravec:2013pua}.
We can and will introduce the background gauge field $\rho$ for this $\SL(2,\bZ)$ symmetry, which means that there is a nontrivial duality transformation when going around a nontrivial loop in spacetime.
The phase of the partition function is then 
\begin{equation}
\tfrac1{2\pi}\Arg Z_\text{BdC}(M_5)= -\tfrac14\eta_\text{signature} +  \Arf(q) 
\label{Maxwell_partfunc}
\end{equation}
where the eta invariant is now for the signature operator $*d$ acting on the differential forms tensored with $(\bZ^2)_\rho$,
and $q$ is now the quadratic refinement of the natural torsion  pairing on $H^3(M_5,(\bZ^2)_\rho)$.
Here, $(\bZ^2)_\rho$ signifies the coefficient system twisted by the $\SL(2,\bZ)$ bundle $\rho$.
%\footnote{%
The eta invariant of the signature operator with such a twist and its reduction from higher dimensions were considered earlier in the mathematical literature; see, e.g.,~\cite{BismutCheeger}.%}

Let us first consider the case where we do not have the $\SL(2,\bZ)$ background. 
In this case, the signature eta invariant simply vanishes, and only the Arf invariant contributes.
Recall that a quadratic refinement is simply the classical action evaluated on flat $B$ and $C$. 
Then a general quadratic refinement can be written as
\begin{equation}
\int \frac{B}{2\pi}\frac{dC}{2\pi} + \int \frac{dB}{2\pi} \frac{\mathcal{C}}{2\pi} 
+  \int \frac{\mathcal{B}}{2\pi} \frac{dC}{2\pi},
\label{qr}
\end{equation}
where $\mathcal{B},\mathcal{C}\in H^2(M_5, {\mathbb R}/2\pi \bZ)$ are  the background fields for the electric and magnetic 1-form $\U(1)$ symmetry
of the Maxwell theory~ \cite{Gaiotto:2014kfa}, which we chose to be flat.
Its Arf invariant is computed to be $\int (\mathcal{B}/2\pi) \beta (\mathcal{C}/2\pi)$ where $\beta$ is the Bockstein homomorphism $\beta:H^2(M_5, {\mathbb R}/\bZ)\to H^3(M_5,\bZ)$;
the Bockstein homomorphism $\beta$ can roughly be regarded as the exterior derivative $d$ when it acts on torsion elements of cohomology groups.
The end result is that \begin{equation}
\tfrac1{2\pi}\Arg Z_\text{BdC}(M_5) = \int (\mathcal{B}/2\pi) \beta (\mathcal{C}/2\pi).\label{BdC}
\end{equation}

This reproduces a known result. 
Indeed, the mixed anomaly is known to be of the form $2\pi i \int_{M_5} (\mathcal{B}/2\pi) d (\mathcal{C}/2\pi)$, for which the
mathematically precise formulation~\cite{CS} reduces to Eq.~\eqref{BdC} when we only consider flat fields. 
Furthermore, we can take $\mathcal{B}/2\pi=\mathcal{C}/2\pi=w_2$, where $w_2$ is the Stiefel-Whitney class of the spacetime and regarded as an element of $H^2(M_5, {\mathbb R}/\bZ)$ by using $\bZ_2 \to {\mathbb R}/\bZ$.
The Maxwell theory with this coupling is also known as the all-fermion electrodynamics,
and it has the gravitational anomaly $2\pi i \int w_2\beta w_2 = \pi i \int w_2w_3$ \cite{Thorngren:2014pza,Wang:2018qoy}.

\begin{table}[t]
\begin{tabular}{c|cccc}
&$S^5/\bZ_2$ & $S^5/\bZ_3$ & $S^5/\bZ_4$  &  $S^5/\bZ_6$ \\
\hline
$\eta_\text{signature}$ & $0$ & $-\frac19$ & $-\frac12$ & $-\frac{14}9$ \\
 $H^3(M_5,(\bZ^2)_\rho)$ & $(\bZ_2)^2$ & $\bZ_3$ & $\bZ_2$ & $\bZ_1$ \\
 $\Arf(q)$ & $+\frac12$ & $-\frac14$ & $+\frac18$ & $0$ \\
 $\tfrac{1}{2\pi}\Arg Z$& $+\frac12$ & $-\frac29$ & $+\frac14$ & $+\frac7{18}$\\
 $\eta_\text{fermion}$ & $-\frac1{16}$ & $-\frac19$ & $-\frac5{32}$ & $-\frac{35}{144}$
\end{tabular}
\caption{Partition functions and related data on $S^5/\bZ_k$.
\label{table}}
\end{table}

Let us next consider the case when a nontrivial $\SL(2, \bZ)$ background is present. 
We can choose the symmetry structure on $M_5$ to consider,
such as $\text{spin}\times\SL(2, \bZ)$ or spin-$\Mp(2,\bZ)$ [$:=\text{spin}{\times_{\mathbb{Z}_2}}\Mp(2, \bZ)$], 
distinguished by whether $C^2=+1$ or $=(-1)^F$.
Here, $C\in \SL(2,\bZ)$ is the charge conjugation   $C:(\mathbf{E},\mathbf{B}) \mapsto -(\mathbf{E},\mathbf{B})$, and
the metaplectic group $\Mp(2,\bZ)$ is the double cover of the group $\SL(2,\bZ)$. 
We will focus on the latter case in this Letter, as it has a natural connection to the$(6{+}1)$-dimensional $CdC$ theory on a spin 7-manifold as we will see.
Canonical examples of $M_5$ associated with this symmetry structure are $S^5/\bZ_k$, $k=2,3,4$, and $6$,
where going around the generator of $\pi_1(S^5/\bZ_k)=\bZ_k$ comes with the duality action by an element $g$ of order $k$ in  $\SL(2,\bZ)$.
While $S^5/\bZ_k$ is not spin for even $k$, it has a natural spin-$\bZ_{2k}$ structure for any $k$ by embedding $S^5/\bZ_k \subset \bC^3/\bZ_k$.
Then, we get the spin-$\Mp(2,\bZ)$ structure by embedding $\bZ_{2k} \subset \Mp(2,\bZ)$.
The results of explicit computations for Eq.~\eqref{Maxwell_partfunc} are tabulated in Table~\ref{table}.
When there are multiple choices for $g$ or $q$, we choose a particular one.
Other quadratic refinements correspond to different background fields $(\mathcal{B}, \mathcal{C})$ for electromagnetic 1-form symmetries.

When $k=2$, the relevant element in $\SL(2,\bZ)$ is just the charge conjugation symmetry  $C$.
This case has the anomaly $ \tfrac{1}{2\pi}\Arg Z = 1/2$ on $S^5/\bZ_2$. 
This is responsible for the difference $1/2$ of the Ramond-Ramond (RR) charges of the
O$3^+$-plane and O$3^-$-plane in Type-IIB string theory \cite{Witten:1998xy}. 
As explained in \cite{Tachikawa:2018njr}, for the consistency of the theory,
the fractional part of the RR charge must be exactly negative of the anomaly of a D3-brane living on $S^5/\bZ_2$.
%The difference between the O$3^+$ charge and the O$3^-$ charge is $1/2$. 
The background $(\mathcal{B}, \mathcal{C})$ produced by O$3^\pm$ is such that only the O$3^-$ leads to the anomaly of the Maxwell theory,
explaining the difference of the RR charges;
we note that the charge $1/4$ of the O$3^+$-plane
was already explained by the fermion anomaly~\cite{Tachikawa:2018njr}. 
We can also check that the resulting $\frac1{2\pi} \Arg Z$ for other  $k$ is exactly what is necessary to reproduce the RR charge of the $\mathcal{N}{=}3$ S-fold  \cite{Garcia-Etxebarria:2015wns,Aharony:2016kai}.

Let us now consider the infinitely massive fermions encoding the anomaly of a $(3{+}1)$-dimensional Weyl fermion of unit charge under $\bZ_{2k}$, which was studied in \cite{Garcia-Etxebarria:2018ajm,Hsieh:2018ifc,Guo:2018vij}.
The corresponding eta invariants on $S^5/\bZ_k$ are also tabulated in Table~\ref{table}.
We can check that the relation 
\begin{equation}
 -\tfrac14\eta_\text{signature} +  \Arf(q) = 56\eta_\text{fermion}
 \label{56}
\end{equation} holds for the  choices of the Arf invariants given in Table~\ref{table}.

\section{Why 56?}
Relation \eqref{56} about the anomaly of the Maxwell theory and 56 Weyl fermions in $3{+}1$ dimensions reminds us of the relation  Eq.~\eqref{bosefermi} about the anomaly of a  chiral boson and a chiral fermion in $1{+}1$ dimensions.
In the latter case, the equality should evidently hold because a chiral fermion can be bosonized to a chiral boson in $1{+}1$ dimensions.
It also explained the reason how and why the spin structure could be used to define the quadratic refinement necessary to formulate the integrand of the $\U(1)$ Chern-Simons theory.
In $3{+}1$ dimensions, however, the Maxwell theory and 56 Weyl fermions are two clearly different theories. 
What is the relation? 
How and why does the spin (or, more precisely, the spin-$\bZ_{2k}$) structure provide the necessary quadratic refinement?
One explanation is provided, somewhat surprisingly, by supersymmetric physics in $5{+}1$ dimensions. 
%thanks to the recent better understanding of the anomalies there.

Consider a self-dual tensor field in $5{+}1$ dimensions.
Its dimensional reduction on $T^2$ gives rise to the Maxwell theory in $3{+}1$ dimensions,
geometrizing the $\SL(2,\bZ)$ duality symmetry of the Maxwell theory.
Correspondingly, the $(4{+}1)$-dimensional $BdC$ theory on $M_5$ coupled to an $\SL(2,\bZ)$ bundle
is the dimensional reduction of the $(6{+}1)$-dimensional $CdC$ theory on $M_7$, which is the $T^2$ bundle over $M_5$.

We now embed this theory of a self-dual tensor field into the tensor branch of the E-string theory \cite{Ganor:1996mu,Seiberg:1996vs}.
The E-string theory is a $(5{+}1)$-dimensional theory realized in M-theory, with two continuous families of vacua.
 One family of vacua is called the tensor branch, which describes an M5-brane close to the spacetime boundary of M-theory carrying the $E_8$ gauge symmetry \cite{Horava:1995qa,Horava:1996ma}.
 On the tensor branch, the low energy theory consists of the self-dual tensor field with some additional fields.
The other family of vacua is called the Higgs branch, which describes an instanton of the $E_8$ gauge symmetry.
The instanton breaks $E_8$ to $E_7$, producing some chiral fermions as zero modes of the instanton.
A nontrivial fact in M-theory is that these two families of vacua are continuously connected; an M5-brane put on the spacetime boundary can become an $E_8$ instanton.
The transition point is a strongly coupled conformal field theory.

We start from the M5-brane close to the spacetime boundary and transform it into an $E_8$ instanton of nonzero size. 
In this process, the low energy theory is changed from that of the tensor branch, containing the self-dual tensor field, into that of the Higgs branch, containing $28=56/2$ chiral fermions in $5{+}1$ dimensions
transforming under the fundamental 56-dimensional representation of $E_7$.
Because this is a continuous process, the anomaly at the start and the anomaly at the end should be the same;
previously the same argument was used to compute the anomaly polynomial of the E-string theory in \cite{Shimizu:2017kzs} (which reproduced earlier results
in \cite{Ohmori:2014pca,Ohmori:2014kda,Intriligator:2014eaa}),
but the same statement is true even for the subtler anomalies we are discussing now.
There are also some additional fields on the tensor and Higgs branches, but their anomalies are manifestly the same on both sides, so we can match the anomaly of the self-dual tensor field and the $28$ chiral fermions.

\begin{figure}
\centering
\includegraphics[width=.45\textwidth]{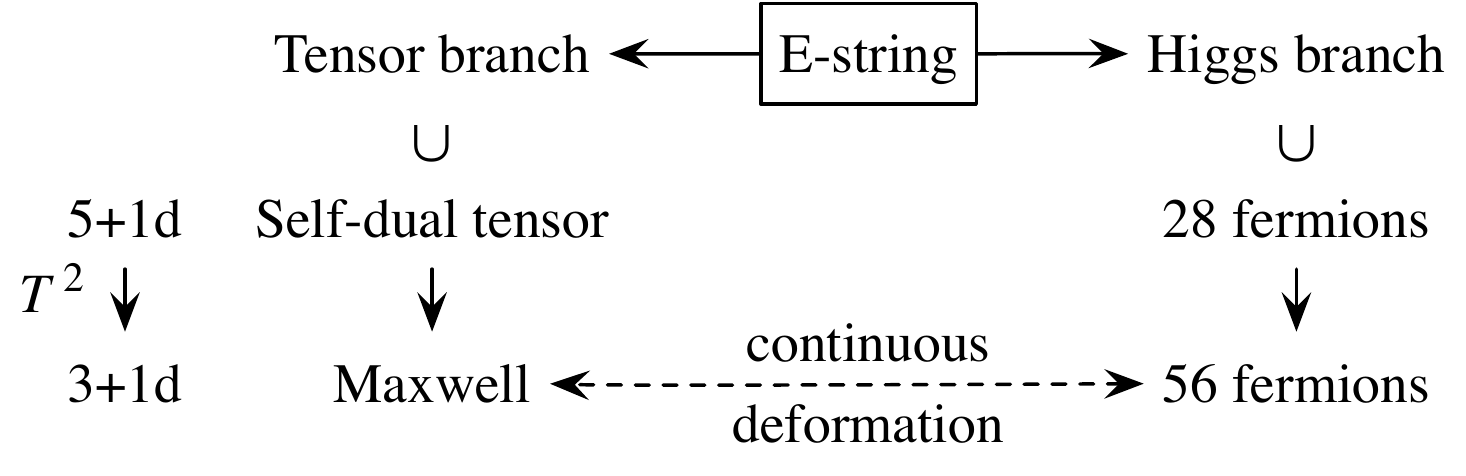}
\caption{Maxwell to 56 fermions via E-string theory \label{fig}}
\end{figure}

Since one chiral fermion in $5{+}1$ dimensions gives rise to two chiral fermions in $3{+}1$ dimensions,
we conclude that the anomaly of the Maxwell theory,
formulated as the $T^2$ compactification of the $(5{+}1)$-dimensional self-dual field with the trivial $E_8$ background,
 should be equal to that of the 56 Weyl fermions.
See Fig.~\ref{fig} for a summary of what we have described.

If we turn on a nontrivial $E_7$ background $A_{E_7}$ on the fermion side,
the data are translated on the self-dual tensor side into the background 3-form field $\mathcal{C}$ which couples to the dynamical self-dual tensor field,
which is basically given by the Chern-Simons term constructed from $A_{E_7}$.
When $A_{E_7}$ is flat, this determines a quadratic refinement required to define the 6+1d $CdC$ theory.
In particular, the trivial $E_7$ background, which is available on any manifold,
provides a canonical quadratic refinement for the $(6{+}1)$-dimensional $CdC$ theory, 
and this construction only requires the spin structure.
This point was already essentially made in \cite{Witten:1996md}.

Since this explanation of Eq.~\eqref{56} requires a lot of information from string and M-theory,
it would be of independent interest to check the equality \eqref{56} by a direct analysis in $3{+}1$ and $4{+}1$ dimensions.
To translate the analysis in $5{+}1$ dimensions to the study of the Maxwell theory, 
we need to require that the $T^2$ bundle over $M_5$ specified by the $\SL(2,\bZ)$ background is equipped with a spin structure. This means that the symmetry structure we consider %in 3+1 and 4+1 dimensions 
is a spin-$\Mp(2,\bZ)$ structure. According to the cobordism classification theorem \cite{Kapustin:2014tfa,Kapustin:2014dxa,Freed:2016rqq,Yonekura:2018ufj},
the anomaly of any system with this symmetry is classified by the dual of  
$\Omega^\text{spin-$\Mp(2 ,\bZ)$}_5=\bZ_{9}\oplus \bZ_{32}\oplus \bZ_{2}$,
which is the bordism group for closed 5-manifolds with spin-$\Mp(2,\bZ)$ structures and is generated by $S^5/\bZ_3$, $S^5/\bZ_4$, and $[(S^5/\bZ_4)'+9(S^5/\bZ_4)]$, respectively, 
where $(S^5/\bZ_4)$ and $(S^5/\bZ_4)'$ both have the  spin-$\bZ_8$ structure coming from the embedding $S^5/\bZ_4\subset\bC^3/\bZ_4$ but with different actions of $\bZ_4$ given by $\mathrm{diag}(i,i,i,\pm i)$.
We have not directly determined %exactly 
which quadratic refinement comes from the trivial $E_7$ field, 
but we have checked that, for a suitable choice, we have the equality \eqref{56} for each case,
providing a strong check of our identification of Eq.~\eqref{56}.
%This provides a strong check of our identification \eqref{56}.

\section{Acknowledgments} 
The authors thank Nati Seiberg for useful comments on a draft of this Letter.
Y. T. also thanks Yu Nakayama for pointing out the relevance of papers by Ganor and collaborators 
\cite{Ganor:2008hd,Ganor:2010md,Ganor:2012mu,Ganor:2014pha} 
to the authors.
C.-T. H. and Y. T. are in part supported  by WPI Initiative, MEXT, Japan at IPMU, the University of Tokyo.
C.-T. H. is also supported in part by JSPS KAKENHI Grant-in-Aid (Early-Career Scientists) No. 19K14608.
Y. T. is also supported in part by JSPS KAKENHI Grant-in-Aid (Wakate-A) No. 17H04837 
and JSPS KAKENHI Grant-in-Aid (Kiban-S) No. 16H06335.
K. Y. is supported by JSPS KAKENHI Grant-in-Aid (Wakate-B) No. 17K14265.

\bibliographystyle{ytphys}
\bibliography{ref}

\end{document}